\documentclass[pra,onecolumn,showpacs,superscriptaddress,amssymb,10pt]{revtex4}

\usepackage{graphicx}
\usepackage{dcolumn}
\usepackage{bm}
\usepackage{epsfig}
\usepackage{color}
\usepackage{longtable}
\usepackage{amsmath}
\usepackage{multirow}
\usepackage{tabularx}
\usepackage{siunitx}
\usepackage{svg}
\usepackage{subfigure}
\usepackage{mwe}
\usepackage{graphicx}
\usepackage{color}
\usepackage{ulem}
\usepackage{MnSymbol}
\usepackage{physics}
\usepackage{bm}
\usepackage{amsmath}
\definecolor{aogreen}{rgb}{0.0, 0.5, 0.0}

\def\sprm#1#2{  \left\langle #1 \left\vert \right. #2 \right\rangle   }

\def\mem#1#2#3{  \left\langle #1 \left\vert  #2 \right\vert #3 \right\rangle   }

\definecolor{mymainmessagecolor}{RGB}{10,200,10}
\definecolor{revisedcolor}{RGB}{0,100,20}

\begin{document}
\include{Bibliography.bib}
\preprint{}
\title{
Photoelectron Angular Distributions of Nonresonant Two-Photon Atomic Ionization Near Nonlinear Cooper Minima
}

\author{J.~Hofbrucker}
\affiliation{Helmholtz-Institut Jena, Fr\"o{}belstieg 3, D-07743 Jena, Germany}
\affiliation{Theoretisch-Physikalisches Institut, Friedrich-Schiller-Universit\"at Jena, Max-Wien-Platz 1, D-07743
Jena, Germany}%
\affiliation{GSI Helmholtzzentrum f\"ur Schwerionenforschung GmbH, Planckstrasse 1, D-64291 Darmstadt, Germany}

\author{L.~ Eiri}
\affiliation{Helmholtz-Institut Jena, Fr\"o{}belstieg 3, D-07743 Jena, Germany}
\affiliation{GSI Helmholtzzentrum f\"ur Schwerionenforschung GmbH, Planckstrasse 1, D-64291 Darmstadt, Germany}

\author{A.~V.~Volotka}
\affiliation{Helmholtz-Institut Jena, Fr\"o{}belstieg 3, D-07743 Jena, Germany}%
\affiliation{GSI Helmholtzzentrum f\"ur Schwerionenforschung GmbH, Planckstrasse 1, D-64291 Darmstadt, Germany}

\author{S.~Fritzsche}
\affiliation{Helmholtz-Institut Jena, Fr\"o{}belstieg 3, D-07743 Jena, Germany}%
\affiliation{Theoretisch-Physikalisches Institut, Friedrich-Schiller-Universit\"at Jena, Max-Wien-Platz 1, D-07743
Jena, Germany}
\affiliation{GSI Helmholtzzentrum f\"ur Schwerionenforschung GmbH, Planckstrasse 1, D-64291 Darmstadt, Germany}

\date{\today \\[0.3cm]}

\begin{abstract}
Photoelectron angular distributions of the two-photon ionization of neutral atoms are theoretically investigated. Numerical calculations of two-photon ionization cross sections and asymmetry parameters are carried out within the independent-particle approximation and relativistic second-order perturbation theory. The dependence of the asymmetry parameters on the polarization and energy of the incident light as well as on the angular momentum properties of the ionized electron are investigated. While dynamic variations of the angular distributions at photon energies near intermediate level resonances are expected, we demonstrate that equally strong variations occur near the nonlinear Cooper minimum. The described phenomena is demonstrated on the example of two-photon ionization of magnesium atom. 
\end{abstract}

\newpage
\maketitle


\section{Introduction}

Over the last few decades, studies of nonlinear light-matter interaction have received much attention, both theoretically \cite{Nikolopoulus:PRA:2006, Rohringer:PRA:2007, Florescu:Phys.Rev:2012, Lagutin:Phys.Rev:2017, Hofbrucker:PRL:2018, Wang:PRA:2019, Gryzlova:PRA:2019, Boll:PRA:2020, Venzke:JPB:2020} as well as experimentally \cite{Dodhy:PRL:1985, Meyer:PRL:2008, Ishikawa:J.ModernOptics:2010, Richardson:JPB:2012, Ilchen:Nat.Commun:2018}. With the advancement of extreme ultraviolet (EUV) and x-ray light sources such as free-electron lasers (FELs) \cite{Shintake:Nat.Photonics:2008, Emma:Nat.Photonics:2010, Pabst:Eur.Phys:2013}, energy restrictions of optical lasers have been removed and new opportunities to investigate ionization of inner-shell electrons of atoms and molecules arose \cite{Pellegrini:RMP:2016}. More importantly, FELs are capable of generating high-brilliance x-ray beams which enables one to explore the inner-shell electron dynamics \cite{Tamasaku:PRL:2013} but also the multi-photon ionization \cite{Meyer:PRL:2008, Tamasaku:PRL:2014}. Modern FEL facilities are even capable to tune the polarization of the intense high-energy beams, which open novel experimental possibilities to investigate the dichroic nature of multi-photon ionization as well as molecular chirality at xuv and x-ray photon energies \cite{Lutman:Nat.Photonics:2016, Ilchen:PRL:2017}.

Two-photon ionization is one of the most fundamental nonlinear (multi-photon) processes, in which two photons are simultaneously absorbed by a target and an electron is emitted. Two-photon ionization can be studied by either detecting the yields of the emitted photoelectrons \cite{ Richardson:PRL:2010, R.Ma:J.Phys:2013}, or produced photoions \cite{Richter:J.Phys:2009} or by collecting the subsequent fluorescence as observed for the two-photon $K$-shell ionization of neutral Ge, Cu, and Co atoms \cite{Tamasaku:Nat.Phot:2014, Szlachetko:Sci.Rep:2016, Tamasaku:PRL:2018, Tyrala:Phys.Rev:2019}. From the measured particle yields, it is possible to extract total two-photon ionization cross sections, which gives rise to the ratio of absolute amplitudes for different ionization pathways. Photoelectron angular distributions, on the other hand, provide information not only about the amplitudes, but also the photoelectron phases. One such experiment was performed at the SACLA FEL \cite{R.Ma:J.Phys:2013}, where the photoelectron angular distributions of two-photon ionization of helium in the vicinity of $1s2p\;\: ^1P$ resonances were measured. 
 
In one-photon ionization, the Cooper minimum describes the photon energy at which the (otherwise) dominant ionization channel passes through a local minimum \cite{Cooper:Phys.Rev:1962}. Such strong change in the ionization amplitudes is not only imprinted in the total cross section, but even more prominently in the photoelectron angular distributions, where it leads to a breakdown of forward-backward symmetry \cite{Ilchen:Nat.Commun:2018,Pradhan:JPB:2011}. What is more, the Cooper minimum can be found also in many-photon (nonlinear) ionization of atoms and appears between any two adjacent level resonances of the same angular momentum \cite{Hofbrucker:Phys.Rev:2019}. We have previously shown, that such a \textit{nonlinear Cooper minimum} enhances the polarization transfer from incident to fluorescence photons \cite{Hofbrucker:Phys.Rev:2019}, and in the case of two-photon $K$-shell ionization, can uniquely reveal multipole contributions in the two-photon ionization process \cite{Hofbrucker:Nat.Commun:2020}. Furthermore, we have demonstrated that the maximum of elliptical dichroism in photoelectron angular distributions appears near the nonlinear Cooper minimum \cite{Hofbrucker:PRL:2018}. Despite the strong effect for elliptically polarized light, the influence of nonlinear Cooper minima on the photoelectron angular distributions of two-photon ionization of atoms has not been studied in detail until today. 

The purpose of this paper is to clearly demonstrate the strong effect of nonlinear Cooper minimum upon photoelectron angular distributions in non-resonant two-photon ionization of atoms. This paper is organized as follows. In Sec.~\ref{Sec.Theory}, we present our theoretical approach based on second-order perturbation theory and independent-particle approximation. In Sec.~\ref{Sec.Results}, we discuss the impacts of the nonlinear Cooper minimum on photoelectron angular distributions in the two-photon ionization and analyze the importance of the incident photon polarization. Finally, conclusions and outlook are given in Sec.~\ref{Sec.Conclusion}.
Relativistic units ($\hbar = c = m = 1$) are used throughout the paper, unless otherwise indicated.

\section{Theoretical background}
\label{Sec.Theory}

We here provide a description of two-photon one-electron ionization of neutral atoms, where an atom is initially in a many-electron state $\left\vert{\alpha_i J_i M_i}\right\rangle$ with total angular momentum $J_i$ and its projection $M_i$, and where $\alpha_i$ refers to all additional quantum numbers that are necessary for a unique characterization of the state. After the simultaneous interaction of the atom with two identical photons $\gamma(\bm{k}, \hat{\varepsilon}_{\lambda})$ which are characterized by wave vector $\bm{k}$ and polarization vector $\hat{\varepsilon}_{\lambda}$, the initial atomic state is excited into a final state. Now, the system contains a singly charged ion $\left\vert{\alpha_f J_f M_f}\right\rangle$ and a photoelectron that can be described by the wave function $\left\vert{\bm{p}_e m_e}\right\rangle$ with asymptotic momentum $\bm{p}_e$ and spin projection $m_e$. The two-photon ionization process can be schematically expressed as 
\begin{eqnarray}
\begin{aligned}
\label{Eq.1}
    \left\vert{\alpha_i J_i M_i}\right\rangle + 2\gamma(\bm{k}, \hat{\varepsilon}_{\lambda}) \rightarrow \left\vert{\alpha_f J_f M_f}\right\rangle + \left\vert{\bm{p}_e m_e}\right\rangle .
    \end{aligned}
\end{eqnarray}
The two-photon ionization process can be described by density matrix theory, which provides simple access to all possible physical observables as well as control over the polarization of incident and outgoing particles. The final state density matrix of the system after the two-photon ionization process, consisting of both the singly ionized atom and the photoelectron is given by
\begin{eqnarray}
\begin{aligned}
\label{Eq.12}
 \mem{\alpha_f J_f M_f, \bm{p}_e m_e}{\hat{\rho}_f}{\alpha_f J_f M_f^\prime, \bm{p}_e m^{\prime}_e} &= \sum_{M_i M_i^\prime}\sum_{\lambda_1 \lambda_2 \lambda_1^{\prime} \lambda_2^{\prime}} \mem{\alpha_i J_i M_i,\bm{k} \lambda_1 \bm{k} \lambda_2}{\hat {\rho}}{\alpha_i J_i M^{\prime}_i,\bm{k} \lambda_1^{\prime} \bm{k} \lambda_2^{\prime}}\\ &\times M^{\lambda_1 \lambda_2}_{J_i M_i J_f M_f m_e} M^{\lambda_1^{\prime} \lambda_2^{\prime}\ast}_{J_i M_i^\prime J_f M_f^\prime m_e^\prime} .
 \end{aligned}
\end{eqnarray}
Since the neutral atom and the two photons are initially in a product state, the incident density matrix can be decomposed as follows
\begin{eqnarray}
\begin{aligned}
\label{Eq.2}
 \mem{\alpha_i J_i M_i,\bm{k} \lambda_1 \bm{k} \lambda_2}{\hat {\rho}}{\alpha_i J_i M^{\prime}_i,\bm{k} \lambda_1^{\prime} \bm{k} \lambda_2^{\prime}} = &  \mem{\alpha_i J_i M_i}{\hat {\rho}_i}{\alpha_i J_i M^{\prime}_i} \mem{\bm{k} \lambda_1}{\hat {\rho}_\gamma}{\bm{k} \lambda_1^{\prime}} \mem{\bm{k} \lambda_2}{\hat {\rho}_\gamma}{\bm{k} \lambda_2^{\prime}} ,
  \end{aligned}
\end{eqnarray}
where the neutral atom is initially taken to be in an unpolarized ground state. Hence, its density matrix simplifies to
\begin{eqnarray}
\begin{aligned}
\label{Eq.3}
 \mem{\alpha_i J_i M_i}{\hat{\rho}_i}{\alpha_i J_i M_i^\prime} = & \frac{1}{[J_i]}\delta_{M_i M_i^\prime} .
 \end{aligned}
\end{eqnarray}
The photon density matrices $\mem{\bm{k} \lambda}{\hat {\rho}_\gamma}{\bm{k} \lambda^{\prime}}$ define the degree and direction of polarization of the incoming light and can be conveniently expressed in the helicity representation via Stokes parameters
\begin{eqnarray}
\label{Eq.4}
\mem{\bm{k} \lambda}{\hat {\rho}_\gamma}{\bm{k} \lambda^{\prime}} = &
\begin{pmatrix}
1+P_3 &  P_1 - iP_2\\
 P_1 + iP_2 & 1-P_3 \\ 
\end{pmatrix} .
\end{eqnarray}
Since both photons arise from the same source, they both have equivalent wave vector $\bm{k}$, energy $\omega$ and degree of polarization as defined by the linear ($P_1, P_2$) and circular ($P_3$) Stokes parameters. The interaction of the two photons with the neutral atom is calculated within second-order perturbation theory. Within this theory, the interaction is described by the transition amplitude $M^{\lambda_1 \lambda_2}_{J_i M_i J_f M_f m_e}$, which takes the form
\begin{eqnarray}
\label{Eq.5}
M^{\lambda_1 \lambda_2}_{J_i M_i J_f M_f m_e} & = & \sum_\nu \frac{\mem{\alpha_f J_f M_f, \bm{p}_e m_e}{\hat{R}(\bm{k},{\hat{\varepsilon}_{\lambda_2}})}{\alpha_\nu J_\nu M_\nu} \mem{\alpha_\nu J_\nu M_\nu}{\hat{R}(\bm{k},{\hat{\varepsilon}_{\lambda_1}})}{\alpha_i J_i M_i}}{E_i +\omega -E_\nu} .
\end{eqnarray}
In the above relation, the summation runs over the complete spectrum of intermediate states $\left\vert{\alpha_\nu J_\nu M_\nu}\right\rangle$. The operator $\hat{R}$ represents the (one-particle) electron-photon interaction operator and can be represented in the second quantization formalism as
\begin{eqnarray}
\begin{aligned}
\label{Eq.6}
 \hat{R}(\bm{k},{\hat{\varepsilon}_{\lambda}}) = & \sum_{l m}  \mem{l}{\bm{\alpha} \cdot {\bm{A}_{\lambda}} (\omega)}{m} a^\dagger_{l}a_{m} ,
 \end{aligned}
\end{eqnarray}
where the wave functions $ \left\vert{l}\right\rangle$ and $ \left\vert{m}\right\rangle$ denote the single-electron initial and final states, and where the electron creation and annihilation operators is represented by $a^\dagger$ and $a$, respectively. Furthermore, $\bm{\alpha}$ indicates the vector of the Dirac matrices and ${\bm{A}_{\lambda} (\omega)}$ is the photon wave function. The transition amplitude~(\ref{Eq.5}) can be further simplified, by using the multipole decomposition of the photon field ${{\bm{A}_\lambda} (\omega)}$ into spherical tensors
\begin{eqnarray}
\begin{aligned}
\label{Eq.7}
{{\bm{A}_\lambda} (\omega)} & = & 4 \pi \sum_{J M p} i^{J-p} [\bm{\hat{\varepsilon}}_{\lambda} \cdot \bm{Y}^{(p)\ast}_{J M} (\hat{\bm{k}})] {\bm{a}_{J M}^ {(p)}} (\bm{r}) .
 \end{aligned}
\end{eqnarray}
In expression (\ref{Eq.7}), the vector spherical harmonics are represented by $\bm{Y}^{(p)}_{J M} (\hat{\bm{k}})$ and the electric and magnetic components of the electromagnetic field are described with index $(p=1)$ and $(p=0)$, respectively. Also, the vector functions ${\bm{a}_{J M}^ {(p)}} (\bm{r})$ are sometimes referred to as multipole potentials. Furthermore, by choosing $\hat{\bm{k}}$ as the quantization axis, the dot product of the polarization vector with the vector spherical harmonics becomes 
\begin{eqnarray}
\begin{aligned}
\label{Eq.8}
{[\bm{\hat{\varepsilon}}_{\lambda} \cdot \bm{Y}^{(p)}_{J M} (\hat{\bm{k}})]} = &\quad \sqrt{\frac{[J]}{8\pi}} (-\lambda)^p \delta_{\lambda M} .
 \end{aligned}
\end{eqnarray}
By making use of the independent-particle approximation and by applying the electron creation and annihilation operators to the initial state, the final state of the system after an ionization process can be written as
\begin{eqnarray}
\begin{aligned}
\label{Eq.10}
\left\vert{\alpha_f J_f M_f, \bm{p}_e m_e}\right\rangle = & \sum_{m_a M}\sprm{j_a -m_a, J_i M}{J_f M_f} (-1)^{j_a - m_a} a^\dagger_{p_e m_e} a_{n_a j_a l_a m_a} \left\vert{\alpha_i J_i M}\right\rangle,
\end{aligned}
\end{eqnarray}
where $\sprm{..,..}{..}$ represents a Clebsch-Gordan coefficient. The annihilation operator $a_{n_a j_a l_a m_a}$ describes the creation of a vacancy with a single (active) electron with quantum numbers $n_a, j_a, l_a, m_a$ in the initial atomic state. Such description of the final state wave function of the system (photoion and photoelectron) enables us to reduce the many-electron transition amplitude into a single electron amplitude, see \cite{Hofbrucker:PRA:2016, Hofbrucker:PRA:2017} for details. Further simplification of the transition amplitude arises from expanding of the continuum electron wave function into its partial waves
\begin{eqnarray}
\begin{aligned}
\label{Eq.9}
    \left\vert{\bm{p}_e m_e}\right\rangle = & \frac{1}{\sqrt{\varepsilon_e \abs{\bm{p}_e}}} \sum_{j m_j} \sum_{l m_l} i^l e^{-i{\Delta_{j l}}} \sprm{l m_l, 1/2 m_e}{j m_j} \left\vert{\varepsilon_e j l m_j}\right\rangle Y^\ast_{l m_l}(\hat{p}_e) ,
     \end{aligned}
\end{eqnarray}
with the photoelectron energy $\varepsilon_e = \sqrt{{\bm{p}_e^2 + m^2}}$, the phase factor $\Delta_{j l}$ and the spherical harmonics $Y^\ast_{l m_l}(\hat{p}_e)$ that particularly depend on the direction of the emitted electron. Applying all the above expansions as well as the Wigner-Eckart theorem to Eq. (\ref{Eq.5}) the transition amplitude can be written for initially closed shell atoms as

\begin{eqnarray}
\begin{aligned}
\label{Eq.11}
M^{\lambda_1 \lambda_2}_{J_i M_i J_f M_f m_e}  = &\quad \sum_{p_1 J_1} \sum_{p_2 J_2} \sum_{n_n j_n l_n m_n} i^{J_1 - p_1 + J_2 - p_2} \sqrt{\frac{[J_1,J_2]}{[j_n,j_a]}} (-\lambda_1)^{p_1} (-\lambda_2)^{p_2} \\ \times&\quad \sum_{j m_j} \sum_{l m_l} (-i)^l e^{i{\Delta_{j l}}} \sprm{l m_l, 1/2 m_e}{j m_j} Y_{l m_l}{(\hat{p}_e)} (-1)^{j - m_j} \\ \times&\quad \sprm{j m_j, J_2 -\lambda_2}{j_n m_n} \sum_{m_a}  \sprm{j_a -m_a, J_i M_i}{J_f M_f}\sprm{j_n m_n, J_1 -\lambda_1}{j_a m_a}\\ \times&\quad \frac{\left\langle{\varepsilon_e j l}\left\Vert{\alpha \cdot \bm{a}^{(p_2)}_{{J_2}}}\right\Vert{n_n j_n l_n}\right\rangle \left\langle{n_n j_n l_n}\left\Vert{\alpha \cdot \bm{a}^{(p_1)}_{{J_1}}}\right\Vert{n_a j_a l_a}\right\rangle}{E_{n_a j_a} + \omega - E_{n_n j_n}} ,
\end{aligned}
\end{eqnarray}
where $[J]=2J+1$. The angle-differential two-photon ionization cross section can be simply obtained from the density matrix of the final state of our system by tracing out all quantum numbers of the photoion as well as spin of the photoelectron
\begin{eqnarray}
\begin{aligned}
\label{Eq.13}
    \frac{d\sigma}{d\Omega}  = & \frac{8\pi ^3 \alpha ^2}{\omega ^2} \sum_{J_f M_f m_e} \mem{\alpha_f J_f M_f, \bm{p}_e m_e}{\hat{\rho}_f}{\alpha_f J_f M_f, \bm{p}_e m_e} \\
    = &\frac{8\pi ^3 \alpha ^2}{\omega^2}\frac{1}{[J_i]} \sum_{\lambda_1 \lambda_2 \lambda_1^{\prime} \lambda_2^{\prime}} \mem{\bm{k} \lambda_1}{\hat {\rho}_\gamma}{\bm{k} \lambda_1^{\prime}} \mem{\bm{k} \lambda_2}{\hat {\rho}_\gamma}{\bm{k} \lambda_2^{\prime}} \sum_{J_f M_f M_i m_e} M^{\lambda_1 \lambda_2}_{J_i M_i J_f M_f m_e} M^{\lambda_1^{\prime} \lambda_2^{\prime}\ast}_{J_i M_i J_f M_f m_e} .
     \end{aligned}
     \end{eqnarray}
The photoelectron angular distributions of the two-photon ionization can be factorized using the asymmetry parameters $\beta_n$, together associated with the mth-order Legendre polynomials $\mathcal{P}_m$ as follows 
\begin{equation}
\label{Eq.14}
\frac{d\sigma}{d\Omega}  =  \frac{\sigma}{4\pi} \sum_{m=1} [1+\beta_m \mathcal{P}_m (\cos{\Phi}))],
\end{equation}
where ${\sigma}$ represents the total (angle-integrated) two-photon ionization cross section. The meaning of the angle $\Phi$ depends on the incident photon polarization. For ionization of atoms by linearly polarized light, $\Phi$ is the angle in the polarization plane, measured from the polarization direction. For ionization of atoms by circularly polarized light the angle represents the angle between photon propagation and electron emission directions. Since we shall analyze below the energy dependence of the photoelectron angular distributions, the asymmetry parameters $\beta_m$ provide a convenient way of presenting this dependence. Moreover, in the scenario considered below, the asymmetry parameters $\beta_2$ and $\beta_4$ are fully sufficient to describe the presented results as the contributions from all higher orders are negligible.

\begin{figure}[t]
    \centering
    \includegraphics[scale=0.4]{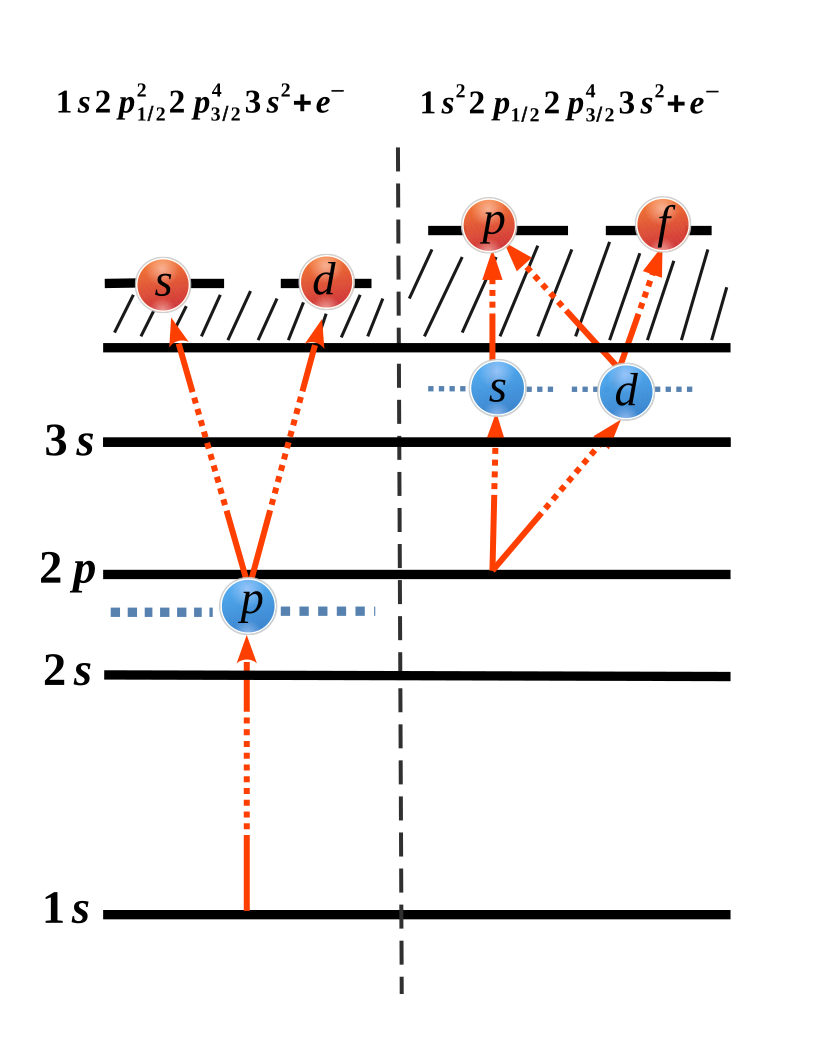}
    \caption{Schematic diagram of possible electric dipole ionization pathways for two-photon ionization of a $1s$ (left) and $2p_{1/2}$ (right) electrons of neutral magnesium. The possible electric dipole ionization pathways in the single-active-electron picture are represented by red arrows. At the nonlinear Cooper minimum, the ionization pathways with highest angular momentum (right-most pathways) have zero contribution to the process. This has a significant impact upon all physical observables associated with the process.}
    \label{fig:diagram}
\end{figure}

\section{Results and discussions}
\label{Sec.Results}

\begin{figure}[t]
    \centering
    \includegraphics[scale=0.37]{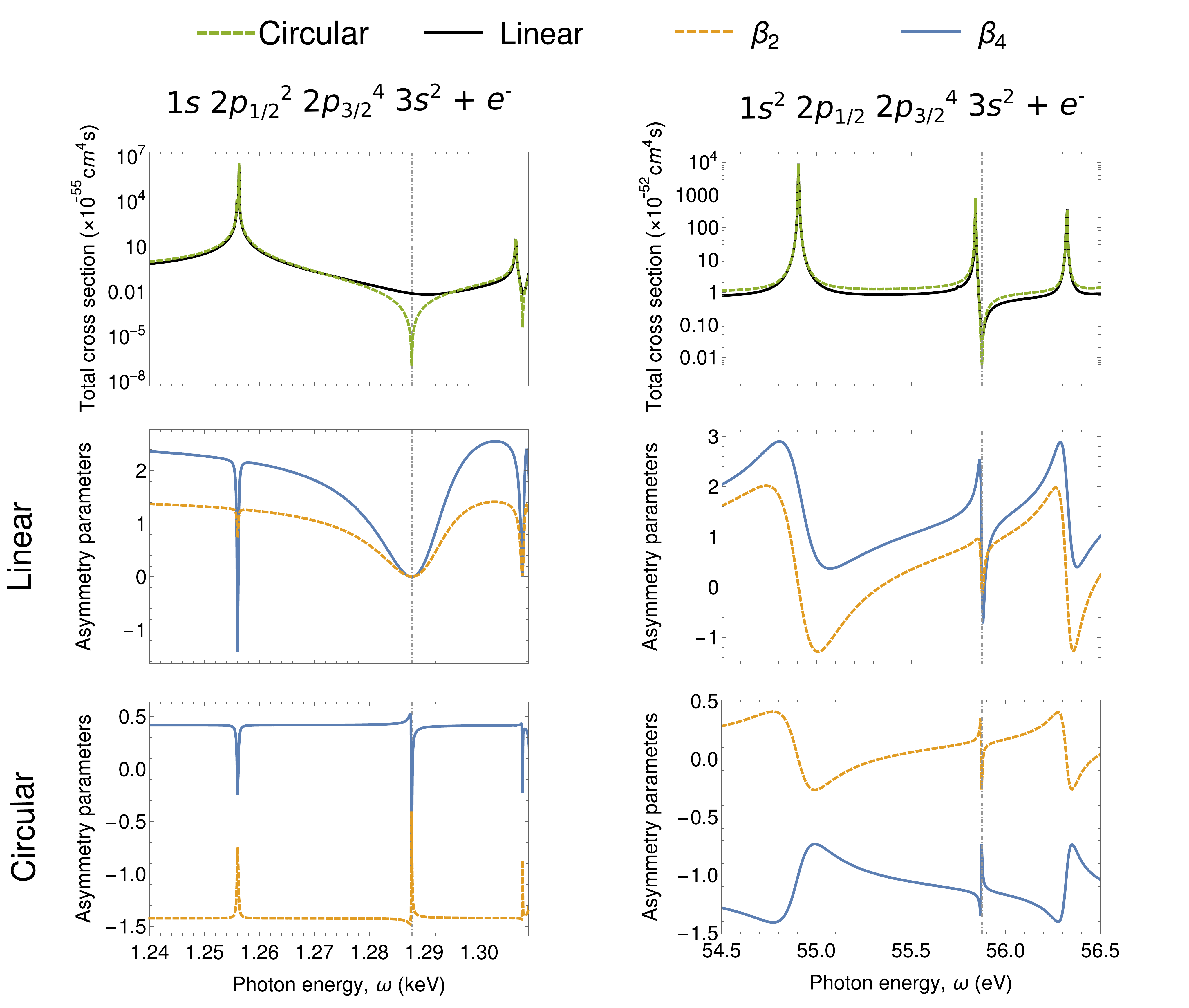}
    \caption{Two-photon ionization of $1s$ (left) and $2p_{1/2}$ (right) electrons of neutral magnesium. Local minima in the total ionization cross section (first row) can be clearly seen. These so called nonlinear Cooper minima arise due to a zero contribution from the otherwise dominant ionization channel. The minima are even more strongly imprinted in the photoelectron angular distributions both for ionization of magnesium by linearly (second row) as well as circularly (last row) polarized light.}
    \label{fig:TPI_Mg}
\end{figure}

Although the formalism presented in the previous section applies generally and is independent of the particular choice of the atom, we here present our results on the example of nonsequential two-photon ionization of magnesium atoms. Furthermore, the photoelectron angular distributions corresponding to the release of an electron from an individual subshell share similar characteristics if their total angular momentum is the same. Here, we shall restrict ourselves to two-photon ionization of the $1s$ and $2p_{1/2}$ electrons only. Further spin contributions were included in the calculations of the presented results, however, they were found negligible. For the sake of simpler future discussion, we will restrict ourselves to the nonrelativistic limit, where the electron spin is ignored. Moreover, we will describe the ionization within single-active electron approximation, and refer to single-electron ionization pathways (see Fig.~\ref{fig:diagram} for schematic representation of electric dipole pathways in two-photon ionization of $1s$ and $2p_{1/2}$ electrons). This ignores coupling of this active electron with other electrons of the atom and replaces the interaction with the other electrons to simply their mean potential. An effective potential is also widely used in calculations of multi-photon ionization using solutions of time-dependent Schr\"odinger equation, which for long pulse lengths are in agreement with time-independent perturbation theory \cite{Gryzlova:PRA:2019}. In our previous work \cite{Hofbrucker:Phys.Rev:2019}, we have evaluated the uncertainty in our calculations due to the choice of the screening potential by using a number of different potential models. There are approaches beyond the independent-particle approximation (see e.g. \cite{Lagutin:Phys.Rev:2017}), however, they are usually applied for ionization by soft XUV photon energies and cannot be applied for deep inner-shell ionization. There are currently no approaches which utilize full many-electron calculation for calculation of two-photon ionization of inner-shell electrons. However, a full many-electron calculation of two-photon ionization process is planned to be integrated in the publicly available atomic structures and processes computing software JAC \cite{Fritzsche:CPC:2019} in near future. 

Our results for two-photon ionization of $1s$ and $2p_{1/2}$ electrons of magnesium are presented in Fig.~\ref{fig:TPI_Mg}. The first row presents the total cross section for two-photon ionization of the $1s$ (left) and $2p_{1/2}$ (right) electrons of magnesium. The total cross sections are presented for incident photon energies in the range of the first corresponding resonances, which can be seen as a local increase of the value of the total cross section. Between each pair of intermediate resonances with total orbital angular momentum of $l_n=l_a + 1$ due to the partial wave of the outgoing electron, there is a local minimum called nonlinear Cooper minimum. Such minimum appears due to zero contribution of the (otherwise) dominant ionization channel and is marked in Fig.~\ref{fig:TPI_Mg} with a vertical dashed line, see Ref. \cite{Hofbrucker:Phys.Rev:2019} for detailed explanation of the minimum. As seen from this figure, the drop in the total cross section is larger for the two-photon ionization by circularly polarized photons. This can be easily understood from the selection rules. For ionization by circularly polarized light, the only electric dipole ionization pathway of the photoelectron is $l_a \rightarrow l_a+1 \rightarrow l_a+2$, which by the Fano's propensity rules \cite{Fano:PRA:1985}, is generally the dominant ionization pathway. Therefore, while electric dipole interaction contributions vanish at the nonlinear Cooper minimum for ionization by circularly polarized light, other electric dipole channels remain in the case of two-photon ionization by linearly polarized light, resulting in significantly higher total cross sections. Similar to the Cooper minimum in one-photon ionization, the nonlinear Cooper minimum has an impact not only on the total cross section, but is even more pronounced in photoelectron angular distributions. 

\begin{figure}[t]
    \centering
    \includegraphics[scale=0.25]{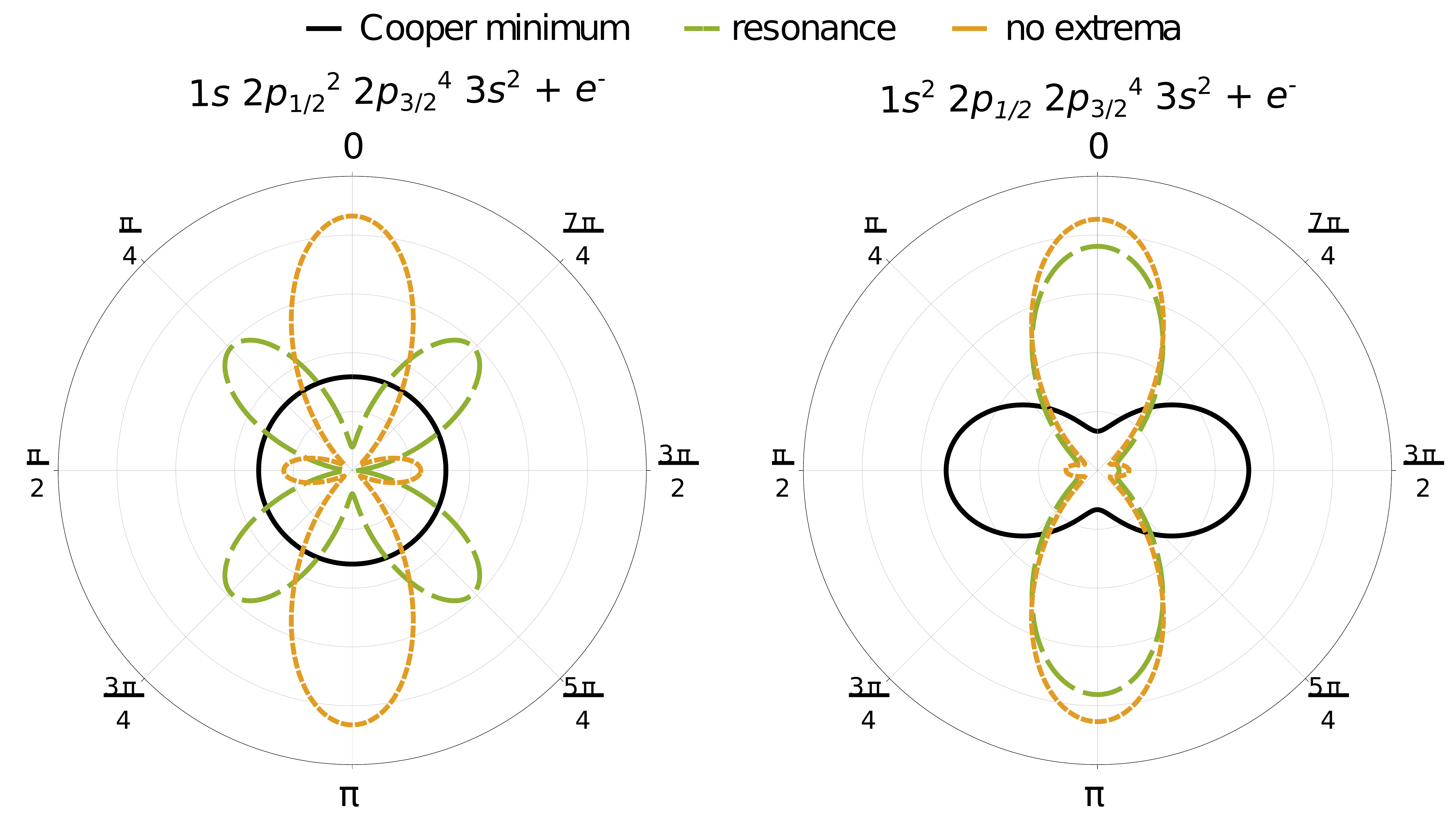}
    \caption{  Photoelectron angular distributions of two-photon ionization of $1s$ (left) and $2p_{1/2}$ electrons of neutral magnesium atoms by linearly polarized light. The distributions were normalized to the total cross section. For ionization of each shell, distributions corresponding to three different photon energies are presented. Incident photon energy matching either the nonlinear Cooper minimum in solid black (1.288~keV and 55.88~eV for ionization of $1s$ and $2p_{1/2}$ electrons, respectively), an intermediate level resonance ($2p_{1/2}$ and $3d$ for ionization of $1s$ and $2p_{1/2}$ electrons, respectively) in long-dashed green (1.256~keV and 55.835~eV) or an energy, which does not match either extrema in short-dashed orange (1.260~keV and 54.50~eV). From the plot, it can be clearly seen that the photoelectron angular distributions at the nonlinear Cooper minima are always distinct.}
    \label{fig:TPI_dist_Mg}
\end{figure}

The angular distributions of two-photon ionization of the $1s$ and $2p$ electrons of neutral magnesium is presented in the second and third rows of Fig.~\ref{fig:TPI_Mg}. The photoelectron angular distributions are described using the asymmetry parameters (see Eq. (\ref{Eq.14})) for ionization by linearly (middle) and circularly (bottom) polarized light, and their dependence on incident photon energy is plotted. Strong variation of the photoelectron angular distributions is predicted both, if the incident photon energies match intermediate level resonances as well as near to the nonlinear Cooper minima. While the influence of intermediate resonances has been discussed before \cite{Lagutin:Phys.Rev:2017, Petrov:Phys.Rev:2019}, the variation at nonlinear Cooper minima have not been discussed until now. We will, therefore, concentrate on the variation of angular distributions at the nonlinear Cooper minima only. Significant differences can be seen for different angular momentum of the created hole as well as polarization of the incident photons. For two-photon ionization of $1s$ electrons by linearly polarized light, two nonrelativistic electric dipole ionization pathways are possible, $s \rightarrow p \rightarrow s$ and $s \rightarrow p \rightarrow d$. Since the the latter pathway vanishes at the nonlinear Cooper minimum, the distribution is predominantly determined by an electron partial wave with spherical symmetry. As a result, both asymmetry parameters drop to zero at the Cooper minimum. For the two-photon ionization of $s$ electrons by circularly polarized light, in contrast, only the $s \rightarrow p \rightarrow d$ pathway is allowed by the selection rules. However, as mentioned before, this pathway vanishes at the nonlinear Cooper minimum, and hence, the distribution at this energy is determined purely by higher multipole order. Similar argumentation can be carried out for the two-photon ionization of $p$ electrons. However, in this case, more pathways contribute to the process which leads to less intuitive explanation. 

Explicit photoelectron distributions of two-photon ionization of $1s$ (left) and $2p_{1/2}$ (right) electrons of neutral magnesium by linearly polarized light are presented in Fig.~\ref{fig:TPI_dist_Mg}. The distributions are presented in the polarization plane for three incident photon energies, one matching the nonlinear Cooper minimum (solid black), one matching an intermediate resonance ($2p_{1/2}$ and $3d$ for ionization of $1s$ and $2p_{1/2}$ electrons, respectively, plotted with long dashed green curve) and one which does not match either of these extrema (short dashed orange). The presented distributions have been normalized to the value of the corresponding total cross section. While the photon energy matching neither extrema represents a typical photoelectron angular distribution, strong variation can be clearly seen at photon energies matching either an intermediate level resonance or the nonlinear Cooper minimum. In the case of two-photon ionization of $1s$ electrons at the Cooper minimum by linearly polarized light, the photoelectron distribution becomes spherically symmetric, as discussed before. For two-photon ionization of $2p_{1/2}$ electrons, on the other hand, the dominant ionization direction is perpendicular to the photon polarization direction. This counter-intuitive results arises from the vanishing contributions of the otherwise dominant ionization pathway and relatively stronger contributions of pathways with lower angular momentum.

Accurate measurement of the energy position of nonlinear Cooper minima would allow to critically evaluate the theoretical representation of the complete electron spectrum of atoms. As can be seen from Fig.~\ref{fig:TPI_Mg}, measurements of the total cross sections could be too insensitive for an accurate determination of the position this minimum, especially since the accuracy of the experimentally determined cross sections suffer higher uncertainties due to beam parameters employed in the experiment. Photoelectron angular distributions, on the other hand, are independent of the intensity of the incident beam and hence, can lead to more precise extraction of the position of the Cooper minimum. Moreover, Fig.~\ref{fig:TPI_Mg} also shows that the distributions vary very dynamically near the nonlinear Cooper minimum of the dominant ionization channel ($l_a \rightarrow l_a+1 \rightarrow l_a+2$). Nonlinear Cooper minima can be found between any pair of intermediate resonances of the same angular momentum, and hence, their detection is not restricted to a particular atom or incident beam energies. It is these properties that make the photoelectron spectroscopy a promising tool for the first determination of Cooper minima in multiphoton ionization. 

\section{Conclusions}
\label{Sec.Conclusion}
The photoelectron angular distributions of two-photon ionization of $1s$ and $2p_{1/2}$ electrons of neutral magnesium have been investigated theoretically. The energy dependence of the angular distribution was presented in terms of asymmetry parameters. In particular, the distributions in the vicinity of nonlinear Cooper minimum are discussed and their strong variation near the minimum is demonstrated. Such a dynamic variation of the angular distributions near the nonlinear Cooper minimum can be utilized, for instance, for determining the position of the minima, as well as for the phase extraction near the minima from experiment, where the different amplitudes have comparable amplitudes.

\end{document}